# Hybrid Deep Learning for Detecting Lung Diseases from X-ray Images

Subrato Bharati, Prajoy Podder, M. Rubaiyat Hossain Mondal

Institute of Information and Communication Technology,

Bangladesh University of Engineering and Technology, Dhaka-1205, Bangladesh.

**Abstract.** Lung disease is common throughout the world. These include chronic obstructive pulmonary disease, pneumonia, asthma, tuberculosis, fibrosis, etc. Timely diagnosis of lung disease is essential. Many image processing and machine learning models have been developed for this purpose. Different forms of existing deep learning techniques including convolutional neural network (CNN), vanilla neural network, visual geometry group based neural network (VGG), and capsule network are applied for lung disease prediction. The basic CNN has poor performance for rotated, tilted, or other abnormal image orientation. Therefore, we propose a new hybrid deep learning framework by combining VGG, data augmentation and spatial transformer network (STN) with CNN. This new hybrid method is termed here as *VGG Data STN* with *CNN* (VDSNet). As implementation tools, Jupyter Notebook, Tensorflow, and Keras are used. The new model is applied to NIH chest X-ray image dataset collected from Kaggle repository. Full and sample versions of the dataset are considered. For both full and sample datasets, VDSNet outperforms existing methods in terms of a number of metrics including precision, recall, F0.5 score and validation accuracy. For the case of full dataset, VDSNet exhibits a validation accuracy of 73%, while vanilla gray, vanilla RGB, hybrid CNN and VGG, and modified capsule network have accuracy values of 67.8%, 69%, 69.5%, 60.5% and 63.8%, respectively. When sample dataset rather than full dataset is used, VDSNet requires much lower training time at the expense of a slightly lower validation accuracy. Hence, the proposed VDSNet framework will simplify the detection of lung disease for experts as well as for doctors.

**Keywords:** Capsule Network, CNN, COVID-19, Vanilla NN, VDSNet, VGG.

*corresponding author: Subrato Bharati
Email address: subratobharati1@gmail.com

## 1. Introduction

The affect of disease on health is rapidly increasing because of alterations to the environment, climate change, lifestyle, and other factors. This has increased the risk of ill health. Approximately 3.4 million people died in 2016 due to chronic obstructive pulmonary disease (COPD), affected generally by pollution and smoking, whereas 400,000 people pass away from asthma [1-2].

The risk of lung diseases is enormous, especially in developing and low middle income countries, where millions of people are facing poverty and air pollution. According to the estimation of WHO, over 4 million premature deaths occur annually from household air pollution-related diseases, including asthma, and pneumonia. Hence, it is necessary to take necessary steps to reduce air pollution and carbon emission. It is also essential to implement efficient diagnostic systems which can assist in detecting lung diseases. Since late December 2019, a novel coronavirus disease 2019 (COVID-19) has been causing serious lung damage and breathing problems. In addition, pneumonia, a form of lung disease can be due to the causative virus of COVID-19 or may be caused by other viral or bacterial infection [3]. Hence, early detection of lung diseases has become more important than ever. Machine learning and deep learning can play a vital role for this purpose. Recently, digital technology has become more important worldwide. This research paper can provide doctors and other researchers a direction for detecting lung disease with the help of deep learning methodology. A large number of lung X-ray images are used as a dataset. The system presented herein can also assist to detect diseases more accurately, which can protect numerous vulnerable people and decrease the disease rate. The health scheme is not yet established due in part to population growth [3, 4].

Many researchers have done investigations to relate machine learning schemes for prediction of X-ray image diagnostic information [5-7]. With the control of computers along with the huge volume of records being unrestricted to the public, this is a high time to resolve this complication. This solution can put up decreasing medical costs with the enlargement of computer science for health and medical science projects. For the implementation, the NIH chest X-ray image dataset is collected from Kaggle repository [8, 9] and it is fully an open source platform. A new hybrid algorithm is introduced in this paper and this algorithm is successfully applied on the above mentioned dataset to classify lung disease. The main contribution of this research is the development of this new hybrid deep learning algorithm suitable for predicting lung disease from X-ray images.

The paper can be organized as follows. Section 2 describes some related works on lung X-ray image classification or lung nodule detection and classification. The problem statement of this research is presented in Section 3. A detailed analysis of the implemented dataset is presented in Section 4. The existing methods for disease classification are discussed in Section 5. The methodology of this research is discussed in Section 6. The results and associated discussion are provided in Section 7, while Section 8 concludes the paper.

## 2. Related works

In spite of launching the first CAD system for detecting lung nodules or affected lung cells in the late 1980s, those efforts were not enough. This is because there were many inadequate computational resources for the implementation of advanced image processing techniques at that time. Lung disease detection using basic image processing techniques is also time consuming. After the successful invention of GPU and CNN, the performance of CAD (for lung disease diagnosing) and decision support arrangement got a high boost. Many studies [10-31] propose many deep learning models in order to detect lung cancer and other lung diseases. The work in [10] focuses on the detection of thorax diseases. A 3D deep CNN is proposed in [11] with multiscale prediction strategies in order to detect the lung nodules from segmented images. However, the work in [11] cannot classify disease types and the multiscale prediction approaches are applied for small nodules. A fully CNN is proposed in [12] for the reduction of false positive rate in classifying the lung nodules. This method can only analyze the nature of the CT scan images in order to reduce the probability of wrong diagnosis. Luna 16 dataset is used in [12]. Faster R-CNN is used in [13] for detecting the affected lung nodules as well as reducing the FP rate. Faster R-CNN shows promising results for object detection. The fusion of deep CNN architecture and dual path network (DPN) is used in [14] for classifying and extracting the feature of the nodules. Multi patches arrangement with Frangi filter is used in [15] to boost the performance of detecting the pulmonary nodule from lung X-ray images. However, their system produces sensitivity of 94% with an FP rate of 15.1.

The significance of artificial intelligence (AI) is offered in [16] with a state of art in the classification of chest X-ray and analysis. Furthermore, the work [16] describes this issue besides organizing a novel 108,948 front outlook database known as ChestX-ray8 where the 32,717 X-ray images are of unique patients. The authors in [16] conduct deep CNNs to validate results on this lung data and so achieve promising results. The database of ChestX-ray8 is also adapted to be used for multi classification of lung diseases [15]. In [24], a framework for deep learning is proposed to predict lung cancer and pneumonia offering two deep learning methods. Initially they use modified AlexNet for diagnosis of chest X-ray. Moreover, in the modified AlexNet, SVM is implemented for the purpose of classification [24]. The authors use LIDC-IDRI and Chest X-ray dataset [24, 25]. Chest X-ray dataset is also used in [26-31]. Comprehensive studies are described in [26] on the detection of consolidation according to DenseNet121 and VGG 16. This system is built on deep learning based computer aided diagnosis [24, 27]. Deep learning based CAD system is used for the clinically significant detection of pulmonary masses/nodules on chest X-ray images [27]. Moreover, deep learning method is also proposed in [28] where several transfer learning methods such as DenseNet121, AlexNet, Inception V3, etc., are used for pneumonia diagnoses. However, the parameter tuning for their implemented methods are very complex. The paper [17] describes that a dataset for big labeled is the point of achievement for classification tasks and prediction. The work in [17] offers a big dataset named CheXpert containing 224,316 radiographic chest images from 65,240 patients. The authors of [17] conduct CNNs to indicate labels to this dataset constructed on the prospect indicated by the model. This model uses lateral and frontal radiographs with observing the output. Moreover, a benchmark dataset is released in [17]. Further the availability of big datasets is extremely anticipated that images with all objects should be recognized lightly and segmentation. Therefore, various methods are needed that can perform both object detection and instance segmentation. Such powerful approaches are FCN and F-RCNN [18-19]. This extended F-RCNN network is known as Mask R-CNN as well as it is superior to F-RCNN according to accuracy and efficiency. The authors of [20] address Mask R-CNN method for segmentation and object detection. The study in [20] compares their algorithm with others and provides the best algorithm from COCO 2016 [21,22]. MixNet (Fusion of two or more networks) is applied in [23] for the detection of lung nodules where GBM is used in classification of two datasets such as LUNA16 and LIDC-IDRI. From the above study, it is clear that research is needed for the detection and classification of lung diseases for the case of large and new datasets.

## 3. Problem Statement

In recent times, a big dataset of X-ray data is available in Kaggle repository [8, 9]. In this paper, this dataset has been implemented using a novel deep learning method by combining CNN, VGG, data augmentation and spatial transformer network (STN). This new hybrid method is termed here as *hybrid CNN VGG Data STN* (VDSNet).

This paper applies the new VDSNet algorithm in analyzing lung disease dataset in order to predict lung disease in patients. For this, a binary classification is conducted using the input attribute of the dataset (such as age, X-ray images, gender, view position) where the output is the detection of diseases indicated by "Yes" or "No". This dataset is very complex and is also a big data, so data processing is difficult. Moreover, it has a lot of noise and it does not have enough information for easily predicting illness. Therefore, processing this dataset is a challenging task.

In this research, patients are classified by using CNN deep learning method on patients X-ray images. Capsule network (CapsNet) [35] can be considered as one of the strongest algorithms having generative and deterministic capabilities. But this network has been comparatively more sensitive to images than the simple CNN structures. CapsNet is capable of squeeze multiple convolutional layers in capsules. After that they are subject to nonlinearity. As CNN models have been popularly used in medical applications, CapsNet has been progressively engaged in some medical related works, for example, brain tumor segmentation and brain tumor classification [36]. As a result, we compare the performance of the new VDSNet method with that of CapsNet. It will be shown in Section 7 that VDSNet outperforms CapsNet, modified CapsNet and other existing deep learning techniques. Hence, the main contribution of this paper is the development of this new algorithm VDSNet which can predict lung disease in X-ray images at an accuracy greater than existing methods.

## 4. Analysis of the chest X-Ray image dataset

This section covers different aspects of the dataset including description, exploration, visualization and view position of the data samples. These are described in the following.

**4.1 Dataset Description**

The sample of dataset [8] file contains a random sample (5%) of the full dataset:

    (i)    It carries 5,606 images where the resolution of each image is 1024×1024

(ii) To create patient data and class labels for the complete dataset such as a comma separated values (.csv) file.

The description of the class are as follows. There are 15 classes (one is "No findings" and another 14 diseases) in the complete dataset, but subsequently this is severely compact version of the complete dataset, various classes are scarce marked as "No findings": Atelectasis-508 images, Pneumonia-62, Hernia-13 images, images, Edema-118 images, Emphysema-127 images, Cardiomegaly-141 images, Fibrosis-84 images, Pneumothorax-271 images, Consolidation-226 images, Pleural Thickening-176 images, Mass 284 images, Effusion - 644 images, Infiltration 967 images, Nodule-313 images, No Finding - 3044 images.

The full dataset [9] contents can be summarized as follows.

(i) It has 12 files accompanied by 112,120 total images with resolution 1024×1024
(ii) To create patient data and class labels for the complete dataset such as a (.csv) file.

The description of the class is as follows. There are 15 classes (one is "No findings" and another14 diseases). Images can be categorized as one or more disease classes as "No findings": Pneumothorax, Consolidation, Infiltration, Emphysema, Atelectasis, Effusion, Fibrosis, Pneumonia, Pleural_thickening, Hernia, Cardiomegaly, Nodule Mass, and Edema.

This paper can contribute in building and analyzing a model based on this valuable dataset. The dataset covers valuable records for the model. In this paper, we will construct it as: age, patient data, gender, snapshot data and X-ray images. For analyzing from X-ray records, doctors can diagnose patient's health and medical conditions. From the output data of X-ray chest images, the intelligent machine can help physicians to diagnose or analyze lung diseases. Some records on gender and age will improve the accuracy of this scheme.

**4.2 Dataset Exploration**

A chest X-ray test is very common and is a cost-effective medical imaging technique. Lung or chest X-ray clinical diagnosis can be of high demand. However, sometimes it may be more problematic than lung diagnosis through computed tomography (CT) imaging for chest. There is a scarcity of resourceful public datasets. Therefore, it is very challenging to realize clinically relevant diagnosis and computer aided detection in various medical sites using chest or lung X-rays. One crucial obstacle in generating big chest X-ray datasets is the absence of properties for labeling numerous images. Before the emancipation of this data, Openi was the biggest in public available in Kaggle where the 4,143 chest or lung X-ray images are available.

The chest X-ray image dataset in [9] consists of 112,120 chest or lung X-ray images using disease labels of 30,805 unique patients. For generating these labels, some authors conducted NLP to text-mine classifications of disease from the related radiological information. These labels are estimated to be greater than 90% accurate as well as appropriate for weakly-supervised learning. Wang et al. [10] localized some common thorax diseases using a small percentage of the dataset. In this data, 5,606 chest images are included with resolution of 1024×1024. Fig. 1 shows two samples X-ray images from the full dataset [9] considered for this study. Fig. 2 presents the percentage of frequency versus diseases from the X-ray images that are within the dataset [9].

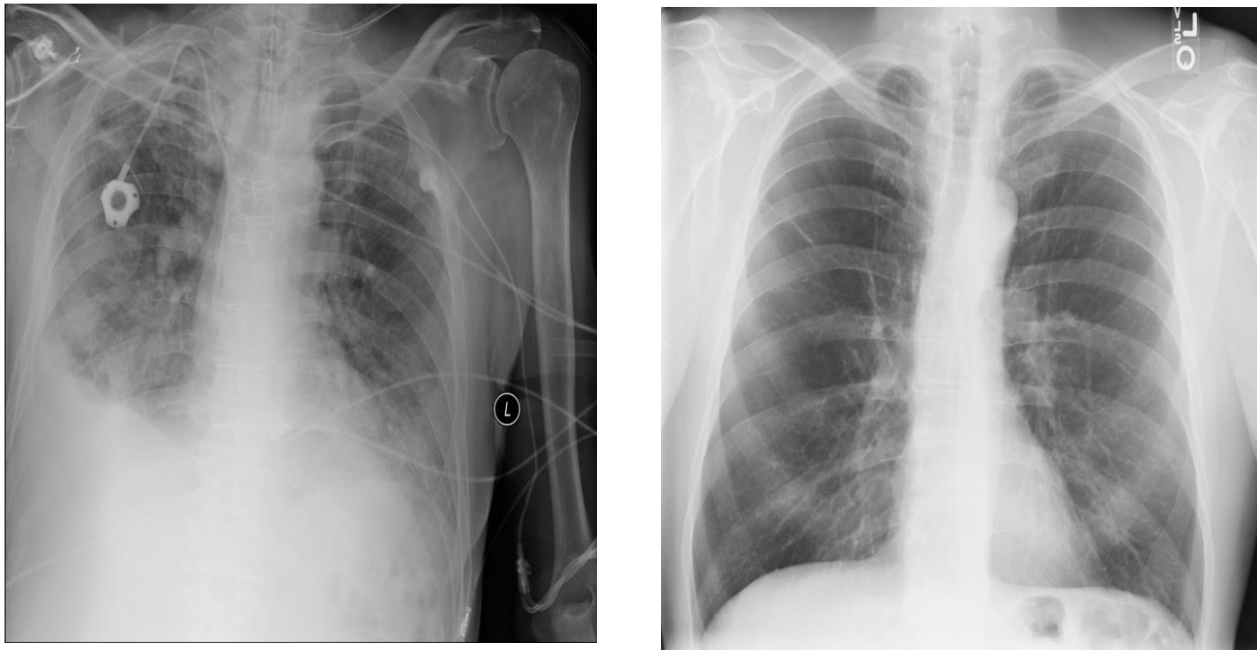

**Fig.1.** Sample of dataset with resolution 1024×1024

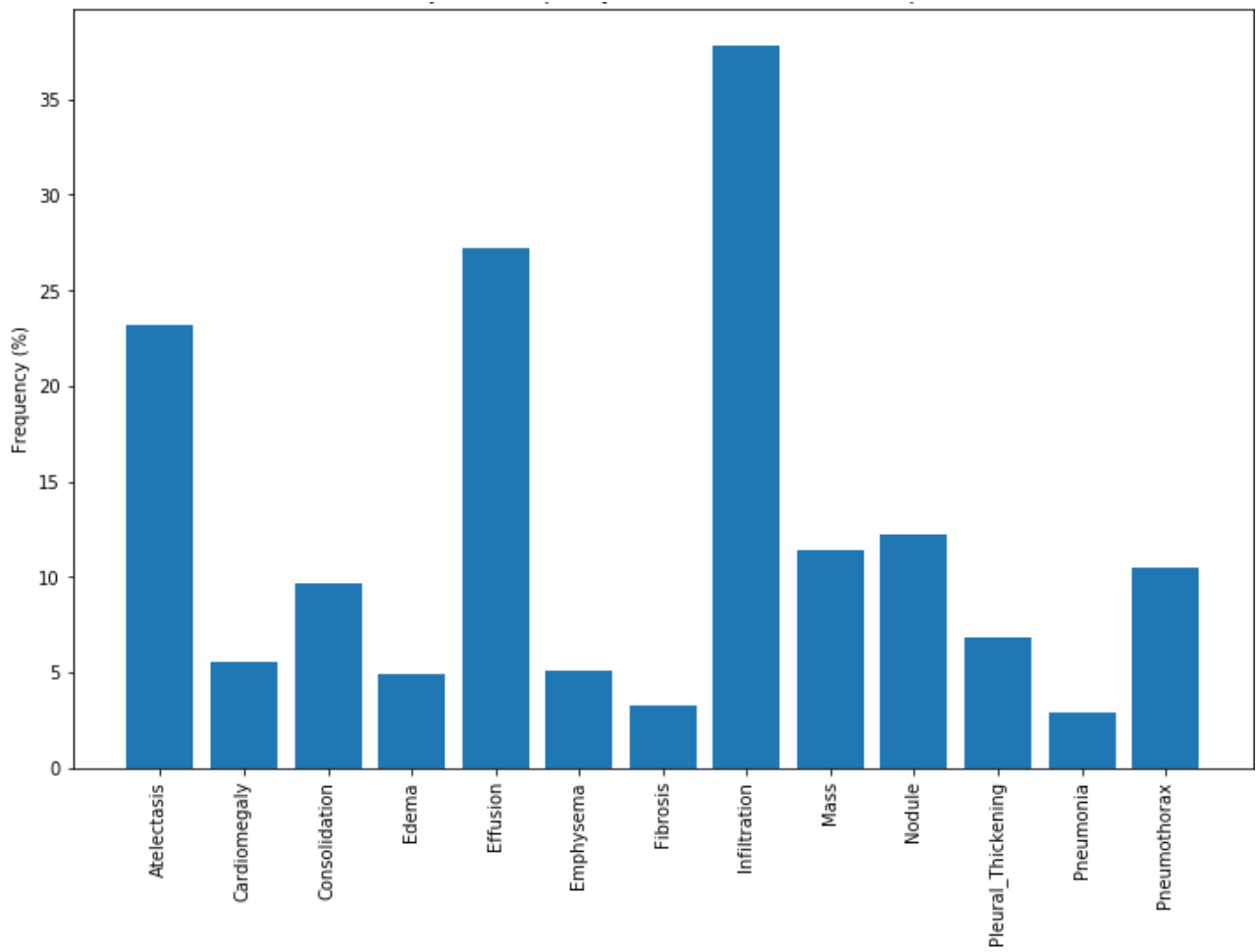

**Fig.2.** Adjusted frequency of diseases in patient group on the image dataset

Patient data and class labels of the total dataset can be illustrated as follows:

- Patient ID
- Finding labels such as disease type
- Image index
- View position: X-ray orientation
- Patient gender
- Patient age
- Original Image Height
- Original Image Width
- Original Image Pixel Spacing_x
- Follow-up
- Original Image Pixel Spacing_y

The data encloses valuable records for the set of data constructed as: gender, age, snapshot data, view position as well as lung X-ray images. We will use this key information in order to train the CNN model.

**4.3 Visualization of the Dataset**

At first, a sample data is analyzed in this study. Finally, full data is analyzed.

In the following, a number of plots will provide some insights about the lung disease data.

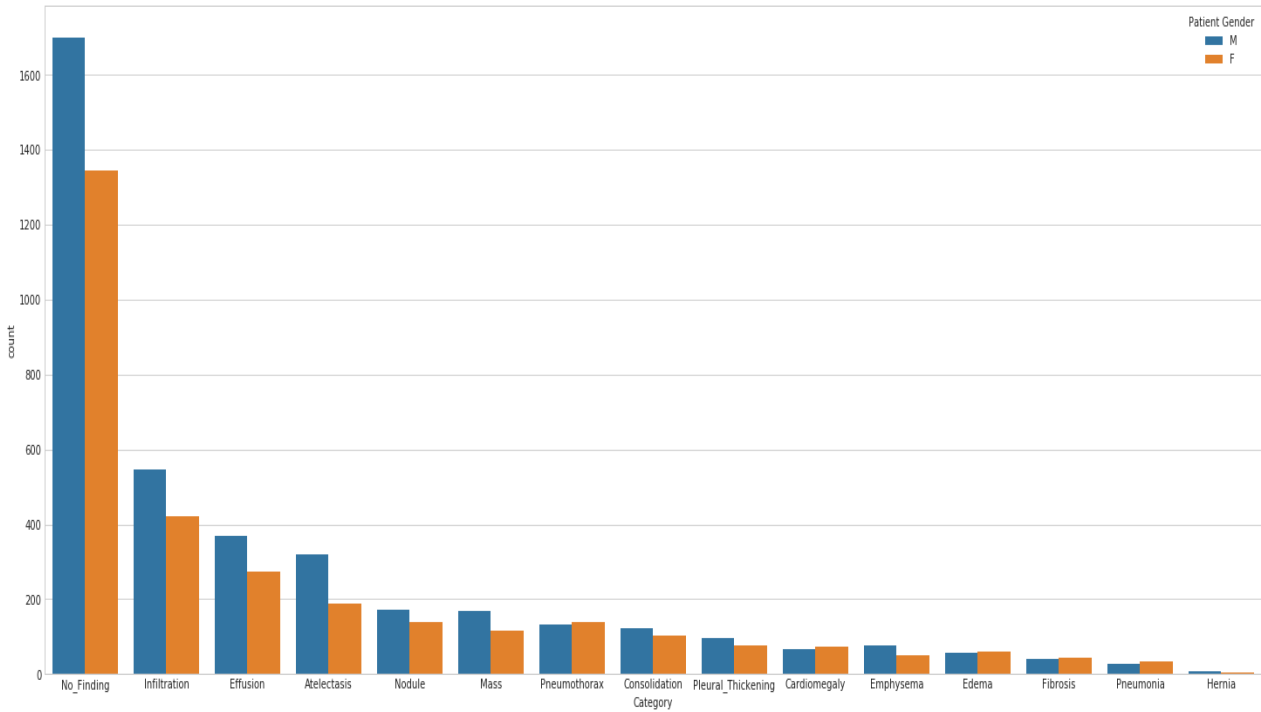

(a)

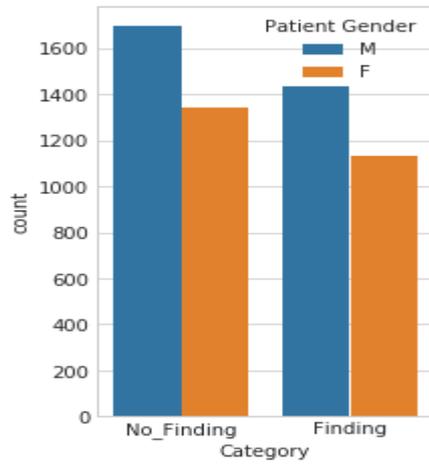

(b)

**Fig. 3.** Visualization of the number of patients in terms of gender and having disease in the sample dataset, (a) for multiclass category (b) for binary category.

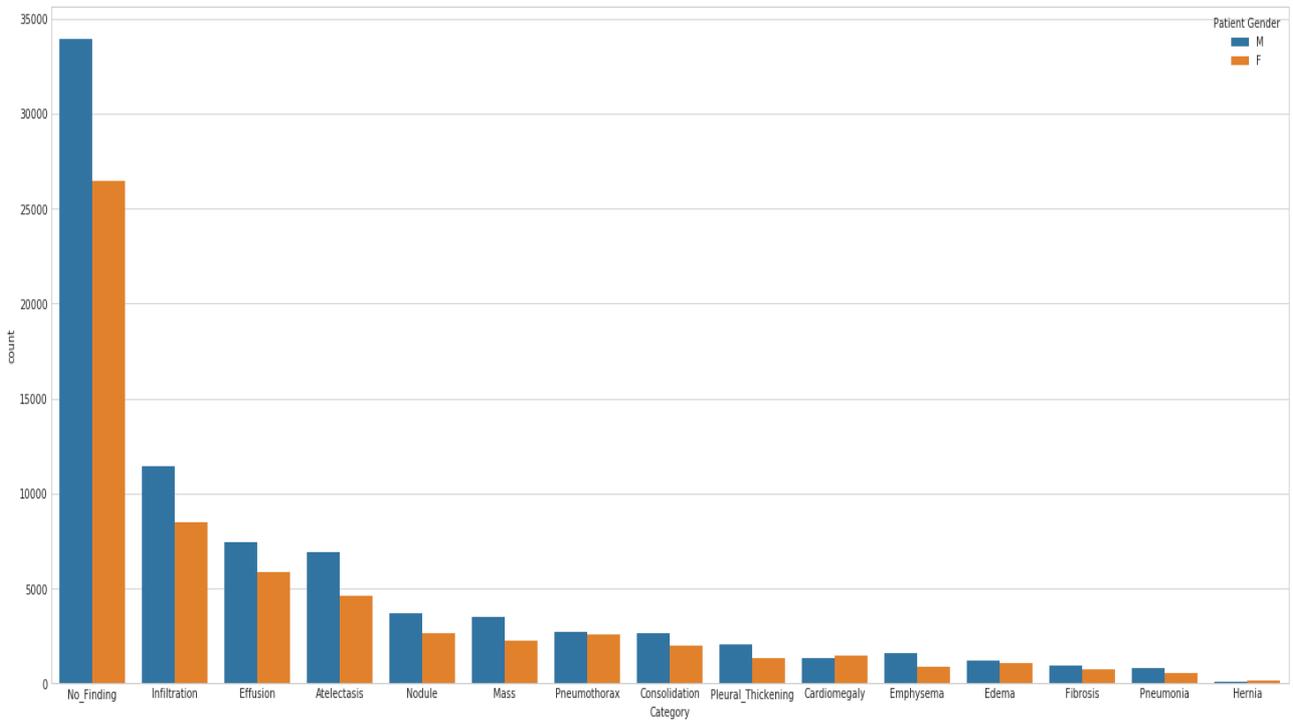

(a)

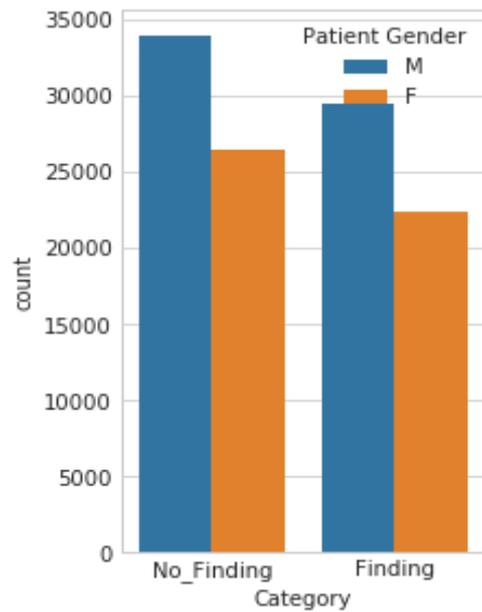

**Fig. 4.** Visualization of the number of patients in terms of sex and having disease in the full dataset, (a) for multiclass category (b) for binary category

Fig. 3 and Fig. 4 show diseases with actual number of cases, for example, Fibrosis, Pneumonia, Hernia, and few many frequent lung diseases for example, Atelectasis, Effusion, Infiltration. Distribution of the diseases is actually uneven. In this dataset, the entire number of males is higher than the entire number of females, and the number of confirmed cases is greater than the number of males diagnosed through lung disease. Fig. 5 and Fig. 6 show the bar diagram of the distribution of patients in two types of view position for the sample and full datasets. The two positions are: anterior-posterior (AP) and posterior-anterior (AP). There are total of 112,120 images in the full dataset.

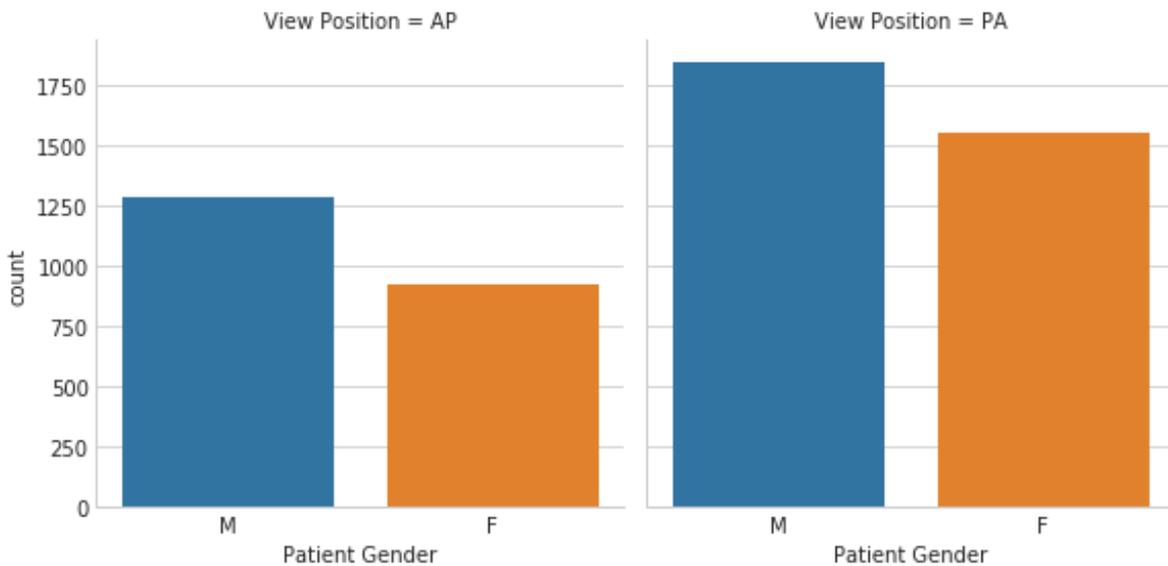

**Fig.5.** Distribution of patients for sample dataset

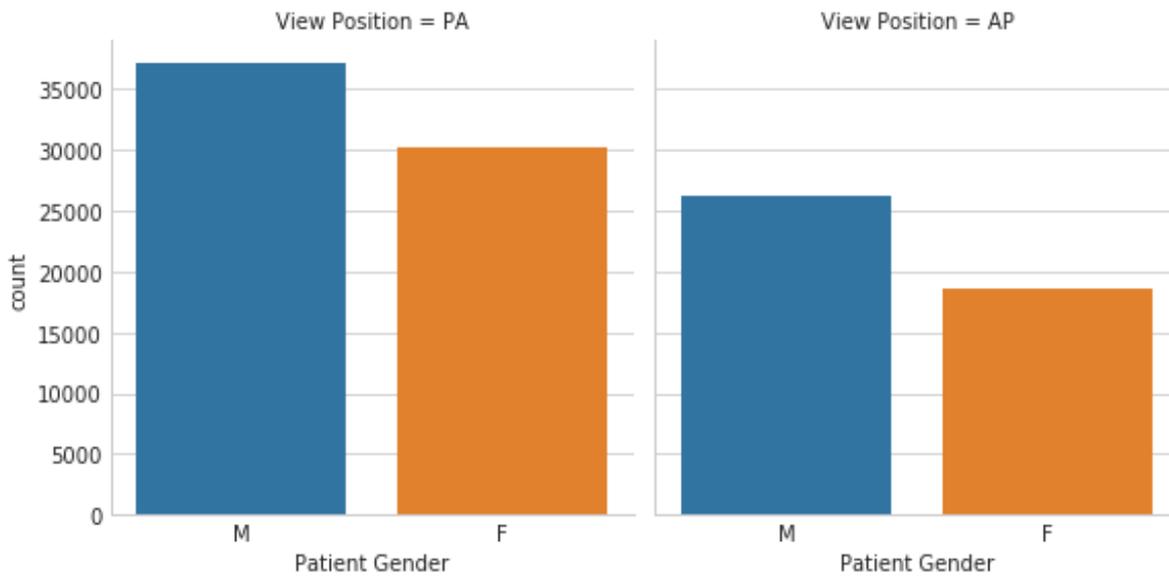

**Fig.6.** Visualization of the amount of patients through view position and sex where full dataset is used

**4.4 View Position**

(i) Posterior-anterior (PA) position: It is a standard position used for finding a regular mature chest radiograph. Patient attitudes standing with the anterior position of chest employed alongside the anterior of the film. The containers are replaced forward adequate to bit the film, confirming in which the scapulae do not make unclear any part of the lung areas. The PA film is observed as if the lung disease patient is fixed in a position.

(ii) Anterior-posterior (AP) position: It is conducted while the patient is immobilized, debilitated, or incapable to collaborate with the PA process. The heart is at a bigger space from the film. Therefore, it seems more expanded than in a PA position. The scapulae are generally visible in the lung fields for the reason that they are not replaced out of the vision in a PA. These types can be realized in which these two categories of position will display the records in the chest X-ray inversely along with the topics specified. As a result, this is moreover an influential feature for the construction of the model. An example from an image having two types of position of the same patient is showed in Fig. 7. The difference can be clearly observed.

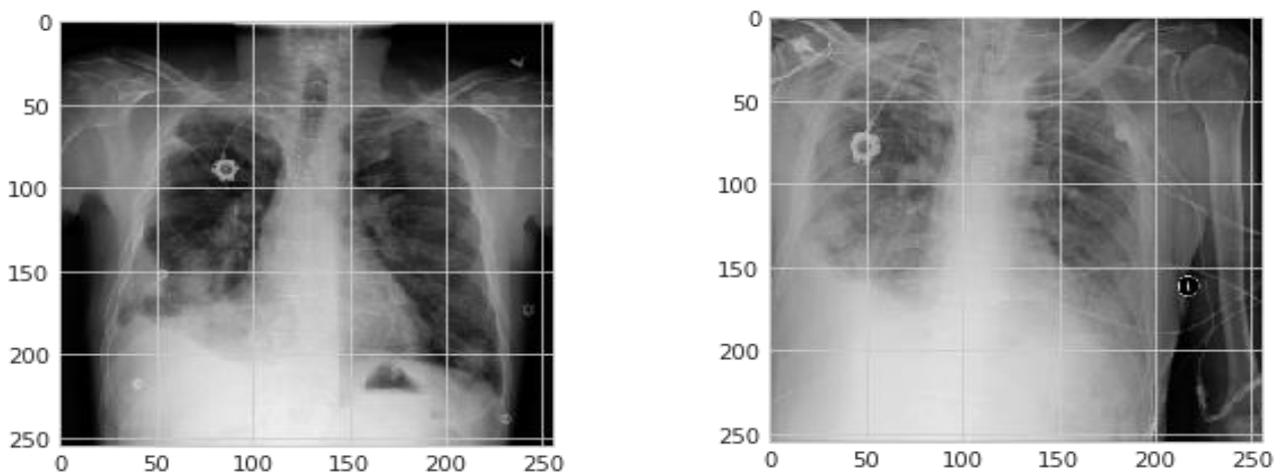

(a)            (b)

**Fig.7.** (a) Posterior-anterior (b) Anterior-posterior position

Fig. 7 shows the view of posterior-anterior and anterior-posterior positions. Compared to young patients, middle-aged patients are more likely to develop lung diseases and thus aim for medical tests. Younger patients are likely to go for primary diagnosis. In order to distinguish whether a person is affected by lung disease or not, some important attributes have been chosen to build the model. The attributes are X-ray, X-ray view position, age, and gender.

## 5. Description of the Existing Methods

In this section, the existing algorithms, CNN and capsule network (CapsNet) are discussed. These two algorithms can combine the important features from not only image data, but also data on age, gender, etc. CNN can be considered as one of the most powerful deep learning based network that can contain multiple hidden layers. These hidden layers are very effective in performing convolution and subsampling for the purpose of extracting low to high levels of features of the input data [32–34]. So, the performance of CNN is evaluated first for this dataset.

CapsNet is proposed by Sabour et al. in 2017 [35]. One of the key features of this network is equivariance which keeps the spatial relationship of objects in an image without affecting the object's orientation and size. CapsNet is also applied in [36] for the classification of brain tumors from brain MRI images. Reliable prediction accuracy and reduced feature map (feature size reduction) are achieved in [36] with CapsNet with changed parameters. CapsNet is also applied in [37] on medical image challenges. A basic CNN with three layers of ConvLayer is selected as the baseline model and the performance of CapsNet is compared with LeNet and the baseline model on four datasets. Their final result shows that CapsNet exhibits better performance than the other two networks for the case of a small and imbalanced dataset [37]. The performance of CapsNet for the case of the large dataset is observed and compared with the other models. The performance capability of basic and modified CapsNet is also evaluated in terms of accuracy and training time calculation. So, a hybrid model is proposed in order to improve the training time and to detect the disease effectively with less number of tests.

CNN has a number of advantages for example, it can extract important features from images at low computational complexity. In this work, a number of aspects of CNN are considered. These are preprocessing parameters which can be sufficient tuning, training parameters, and data enhancement in the system not only lung X-ray images

Using the influence to discriminate several objects from various perspectives, the capsule network can be suitable for the reason that our lung X-ray image data has two categories of view positions. In this paper, the capsules network is modified by tuning the training parameters.

The benchmark model will be a model of vanilla CNN. In this proposed work, "vanilla CNN for sample dataset" and "vanilla CNN for full dataset" have been used. To the best of our knowledge, no researchers constructed a complete deep learning based NN model for this lung X-ray image dataset. Customized mixed link based CNN is used in on LIDC-IDRI dataset for lung nodules detection [38], while STN is used in order to find the optimal model. The architecture or structure of the vanilla CNN model is described in Fig. 8.

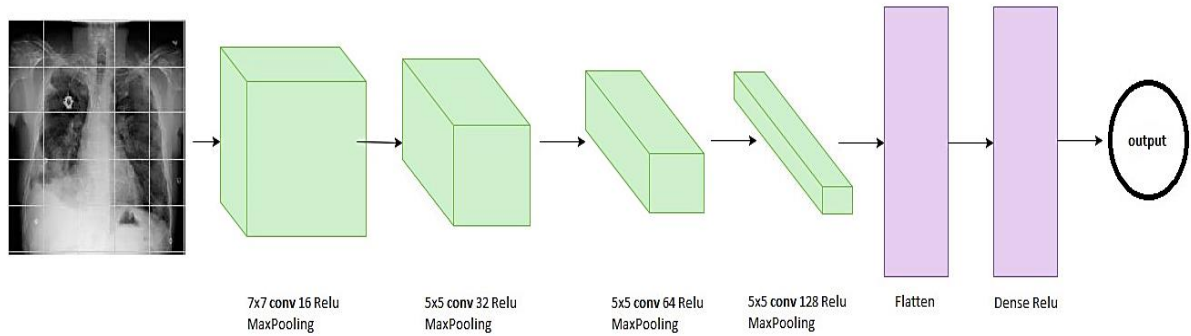

**Fig. 8.** Structural design for the model of vanilla CNN

Fig. 8 shows a model of vanilla CNN where there are four convolutional layers each followed by maximum pooling operation. The convolution layers are growing in depth. Next, is the flattening layer which is followed by a fully connected (FC) dense layer. Finally, the classification output is obtained.

## 6. Methodology

### 6.1 Data Preprocessing

The dataset consists of many X-ray images. Moreover, some additional information such as age or gender distribution can be obtained from the dataset. The preprocessing steps used in this work are mentioned in the following.

- For images:

  (i) At first rescale all images for the purpose of reducing size leading to faster training stage.
  (ii) All the images are transformed to RGB and gray, and are mutually conducted for various models.
  (iii) The numpy array uses for reading the images at that time is normalized by separating the image matrix using 255.

- For additional information:

  (i) Redefine some of the specific features.
  (ii) Normalize the age field to the numeric system then along with the year, at that time normalization field.
  (iii) Eliminate the outliers in the age attribute.
  (iv) There are two essential attributes, this paper will conduct as 'view position' and 'patient gender' in indiscriminate both datasets

All image data when processing is put away for future use. This preprocessing process has the resulting modifiable parameters: resized images form.

### 6.2 Metrics

A number of performance metrics are considered in this research. These are recall, precision as well as $F_\beta$ scores (where $\beta$ is 0.5) designed for binary classification. In this case, F score is superior to accuracy because binary classification is used for detection or finding diseases otherwise the programs are imbalanced. Consider a minor classifier which just predicts the class of majority in an imbalanced dataset. This classifier will achieve a high accuracy when the training size is much greater than the testing size, while the accuracy will be low when the training size is comparable with the testing size. This work considers a number of metrics for the diagnosis of lung diseases. The metrics considered for this work are testing accuracy, precision, recall, and F score [39-40] which can be described with a number of terms including true positive (*TP*), true negative (*TN*), false negative (*FN*) and false positive (*FP*). In the context of this work, *TP* refers to the suspected lung patients that are correctly classified as having lung disease. The terms *TN* is the number of samples having normal condition of the lungs. The term *FN* refers to the suspected patients who actually have lung disease but remains undetected by the system. Moreover, *FP* is the number of patients who are wrongly detected to have lung diseases [40]. The metrics recall and precision can be calculated as follows [40].

$$Recall = \frac{TP}{TP+FN} \tag{1}$$

$$Precision = \frac{TP}{TP+FP} \tag{2}$$

Recall and precision can work on the number of affected patients. So it overcomes the skewness property of the data besides the significance of evaluating a patient's illness. Precision denotes the proportion of patients who properly predict the disease in the entire number of patients who were expected to be ill. Recall denotes the proportion of patients who properly predicts sickness on the entire number of patients truly infected. These parameters can play a significant role in predicting this lung disease. The fusion of precision and recall can be an important metric. The combination of recall and precision known as F score can be described in the following form:

$$F_\beta = (1+\beta^2)\frac{Recall \times Precision}{\beta^2 \cdot Precision + Recall} \tag{3}$$

Various β will display the significance among various precision and recall values. There are two fundamental ideas for selecting the significance of recall and precision:

(i) If the model shows good performance results, then it will be useful for detecting lung diseases in a practical scenario. It is highly significant, since it can be considered a system to support doctors using further diagnostic processes. As a result, low recall and high precision correlated with small β is needed. In this case, β = 0.5 has been assumed for F-β score.

(ii) The proposed models should keep away from mispronouncing sick people in order to avoid illness. Models should avoid missing patients at risk. This situation will prefer high recall and low precision values correlated with large β. In this case, β = 2 has been assumed for F-β score.

The proposed work will help doctors for detecting diseases quickly because in order to determine the disease, a patient needs many tests. The affected patient will be worried before getting additional test results. Therefore, this paper suggests F-0.5 score where β is 0.5.

### 6.3 Implementation of VDSNet

In this work, the algorithms are implemented using Jupyter Notebook, Tensorflow, and Keras. The implementation processes are described below. This is the key scheme of this paper and can be realized on Jupiter Notebook as "VDSNet for sample dataset" and "VDSNet for full dataset". Fig. 9 illustrates the full architecture of VDSNet.

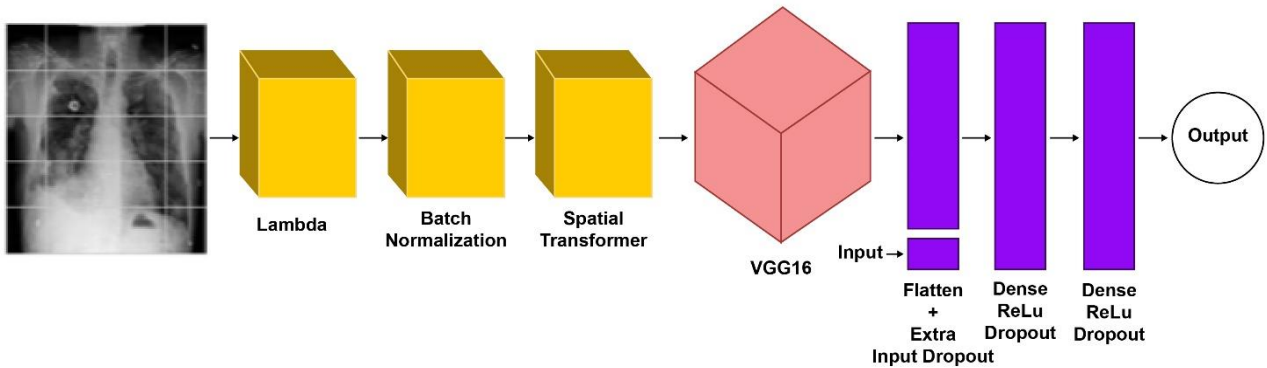

**Fig. 9.** Full architecture of VDSNet

The structure contains three key layers in the ensuing order:

- Spatial transformer layers
    (i) There are three layers.
    (ii) The first part is lambda $\lambda$ to transfer the default routing [-0.5: 0.5], which indicates that the features of the lung X-ray images have a normal value of 0.
    (iii) The second part is batch normalization.
    (iv) The third layer is spatial transformer, which is used to remove the maximum significant features for lung disease classification.

- Extraction of features layers

    (i) VGG16 model has been pre-trained.
    (ii) VGG16 architecture has thirteen convolutional layers, five max pooling layers and three dense layers. So, the summation of total layers is 21, but it has only 16 weight layers.
    (iii) Five models are used on VGG16 as shown in Fig. 10. For example, model 3 consists of eight layers after the convolutional layers. The eight layers are: GAP layer, FC layer having 512 neurons, dropout layer, second FC layer having 256 neurons, second dropout layer, third FC layer having 128 neurons, third dropout layer and a classification layer with a SoftMax activation function. In all the cases, the drop rate of the dropout layer is 50%.

| Model | Configuration |
|---|---|
| 1 | GAPFC(4096) → FC(4096) → Softmax |
| 2 | GAP → Softmax |
| 3 | GAP → FC(512) → Dropout(0.5) → FC(256) → Dropout(0.5) → FC(128) → Dropout(0.5) → Softmax |
| 4 | GAP → FC(512) → Dropout(0.5) → Softmax |
| 5 | GAP → FC(512) → Dropout(0.5)→ FC(512) → Dropout(0.5) → FC(256) → Dropout(0.5) → Softmax |
| **GAP** → global average pooling <br> **FC** → fully connected <br> **Drop rate:** 50% | |

**Fig.10.** Five-layer model of VGG16

- Classification layers

    (i) In this case, the first layer is defined as the flattened layer as of the output of the VGG16 layers with additional 5 features such as 'Gender Female', 'Gender Male', 'Age', 'View position PA', 'View position AP'. These additional 5 features will similarly influence the sorting, such as this simulation has seen upon, therefore they are assembled to the following layer. Accordingly, this layer is called dropout layer.

    (ii) The last two layers are dense dropout layers, with a continuing reduction in depth.

The sequence of steps in this process is described as follows:

(i) Loading of the dataset has been managed into random access memory (RAM) and processing this data as previously where the images are stored in RGB lung X-ray image format.
(ii) Implementing the network structure designed by the way of an architect.
(iii) Implementing the metric function as well as precision score, binary accuracy through threshold, $F_\beta$ score using $\beta$ with a threshold.
(iv) Implementing data model generator, checkpoint, and loss of model function.
(v) Training model using training parameters, validation loss with training/logging training/validation with accuracy.
(vi) Testing the dataset.

CNN and deep learning are employed by Keras where Tensorflow-gpu is used in the backend. By experimenting and changing with numerous image sizes, it is found that the 64×64 image size was good and slight enough for the classifier to the shape of the image capture. The spatial transformer is used and the front layer is supported as $\lambda$ layer. A localization network "locnet" model is used in this STN layer. This helps separating key features from the images. Non-complementary dataset has been tested in various spaces on the structural design. The first layer can be considered the most suitable and pertinent. Adjustment as well as improvement of the thresholds of recall, precision, and $F_\beta$ score are necessary. The index of the dropout layer needs to be refined.

**6.4 Implementation of Modified CapsNet**

In this work, the CapsNet from the main Hinton architecture is modified to make it fit for the lung image dataset [35]. Fig. 11 shows a basic CapsNet architecture for lung X-ray images analysis.

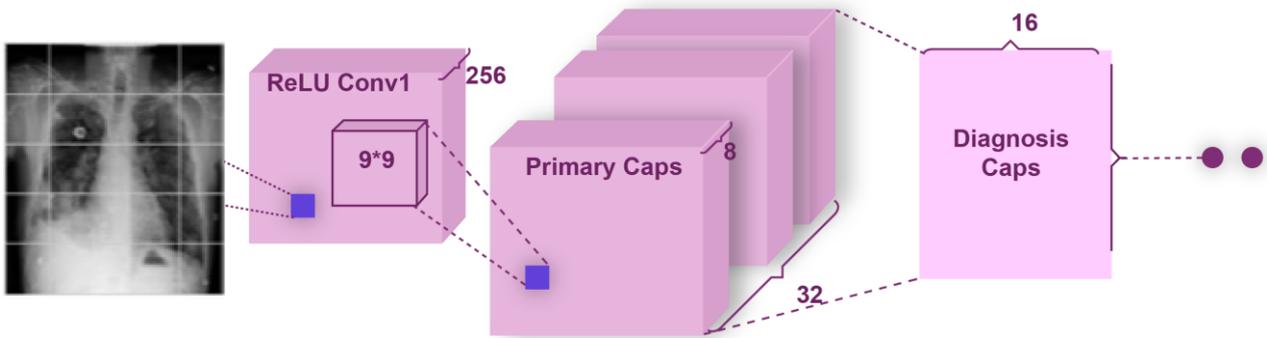

**Fig.11.** Capsule network for lung X-ray images prediction

Main portions of this model can be summarized as follows.

- Convolution layer with filters = 256, strides = 2, kernel_size = 9, activation = 'relu', padding = 'same'. This layer was improved as of the original classifier from strides = 1 to strides = 2, the image was 28×28, as well as the data was 64×64, the output of this classifier will be considerably compacted. With strides=2, we will acquire less features than strides = 1, subsequently we have improved the strings, consequently we consider that the output of lung images have been considerably concentrated.
- Primary capsule with dim_capsule=8, strides=2, kernel_size=9, n_channels=32, padding='same', simply variations with Hinton's structure in which the padding 'valid' is exchanged with 'same'.
- Diagnosis capsule (we change the similar name in which Hinton situates) with n_class=num_capsule, dim_capsule=16, stable of the set routings.

The process of setting the parameters of the capsule network can be described by the following algorithm 1.

*Algorithm 1: Capsule Network Model*

import numpy as np

from keras import layers, models, optimizers

from capsulelayers import CapsuleLayer, PrimaryCap, Length, Mask

def CapsNet(input_shape, n_class, routings):

p= layers.Input(shape=input_shape)

    (i)    Layer 1: A simple Conv2D layer

conv1 = layers.Conv2D (filters=256, kernel_size=9, strides=1, padding='valid', activation='relu', name='conv1')(x)

    (ii)    Layer 2: Conv2D layer with "squash" activation

primarycaps = PrimaryCap (conv1, dim_capsule=8, n_channels=32, kernel_size=9, strides=2, padding='valid')

    (iii)    Layer 3: Capsule layer

DiagnosisCaps = CapsuleLayer (num_capsule=n_class, dim_capsule=16, routings=routings, name='DiagnosisCaps')(primarycaps)

> (iv) Layer 4: Auxiliary layer to replace each capsule with its length
> out_caps = Length(name='capsnet')(DiagnosisCaps)

As like CNN, the application steps are applied in this next step:

- Loading of the dataset has been managed into RAM and processing this data as previously where the images are stored in RGB lung X-ray image format.

- Implementing the network structure designed by the way of an architect considered beyond with the parameters illustrated.

- Implementing of the metric function containing precision score with threshold, binary accuracy, $F_\beta$ score with β and threshold, recall score with threshold. There is a minor modification from CNN to the output form (None, 2) in place of CNN with the output form (None, 1).

- Implementing data model generator, checkpoint, and model loss function.

- Training model using training parameters, validation loss besides training/logging training/validation accuracy.

The parameters selected for capsule network are: convolution layer with filters = 256, strides = 2, kernel_size = 9, activation = 'relu', padding = 'same'. This layer was improved as of the original classifier from strides = 1 to strides = 2, the image was 28×28, the reason creature that with the MNIST data Hinton tested capsule network, as well as the data was 64×64, the output of this classifier will be considerably compacted, with strides=2, as well as we will agree so that we will acquire less features than strides = 1, subsequently we have improved the strings consequently we consider the output of lung images have been considerably concentrated. Therefore, we vary the value of padding from 'valid' to 'same'.

The metric function containing precision score with threshold, binary accuracy, $F_\beta$ score with β and threshold, recall score with threshold are implemented. There is a minor modification from CNN to the output form (None, 2) in place of CNN with the output form (None, 1). The parameters for training are similarly offered to ensemble the machine configurations for example, learning rate, batch size=32.

## 7. Results and Discussion

The performance results of the proposed model and existing models are presented in this section. Some abbreviations used for the models are described in the following.

(i) Vanilla RGB: vanilla CNN model for RGB images
(ii) Vanilla gray: vanilla CNN model for gray images
(iii) Hybrid CNN and VGG: optimized CNN with VGG16 pre trained model
(iv) Hybrid CNN, VGG, and data: VDSNet with VGG16 pre trained model and data augmentation
(v) Hybrid CNN, VGG, data and STN: VDSNet with VGG16 pre trained model, data augmentation and spatial transformer
(vi) Basic CapsNet : Capsule network with Hinton's architecture
(vii) Modified CapsNet: Capsule network with modified architecture

## 7.1 Model Validation and Evaluation

During improvement, a validation set was used to estimate the model. Fig. 12 is a graphical representation of the loss value against epoch. In the figures, 'loss' indicates training loss, while 'val_loss" indicates validation loss. Fig. 12(a) is for the case of vanilla CNN using sample dataset, while Fig. 12(b) is for the case of vanilla CNN using full dataset. Similarly, Fig. 12(c) and Fig. 12(d) are for capsule network for sample and full datasets, respectively. Furthermore, Fig. 12(e) and Fig. 12(f) are for VDSNet for sample and full datasets, respectively.

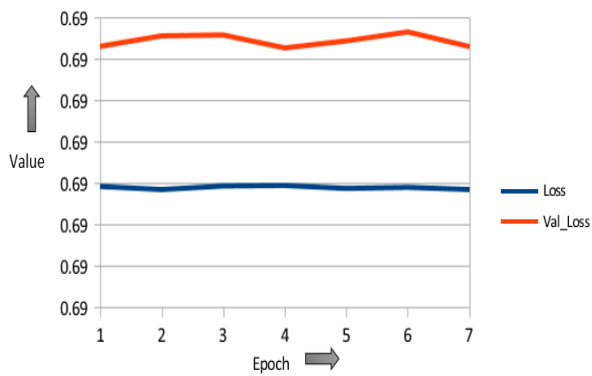
(a) Vanilla CNN for sample dataset

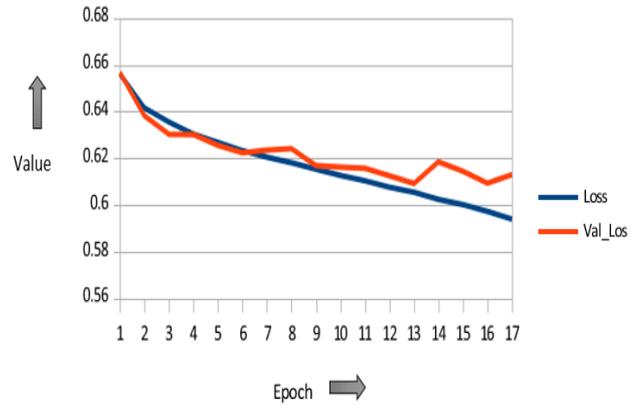
(b) Vanilla CNN for full dataset

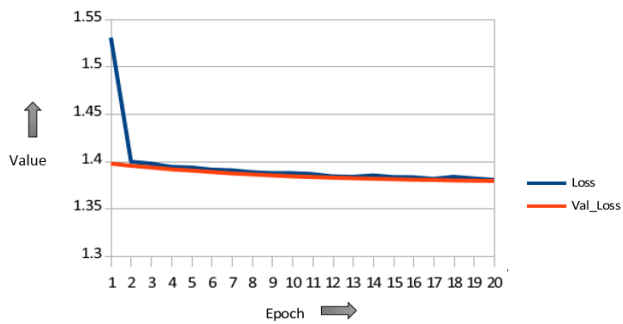
(c) Capsule network for sample dataset

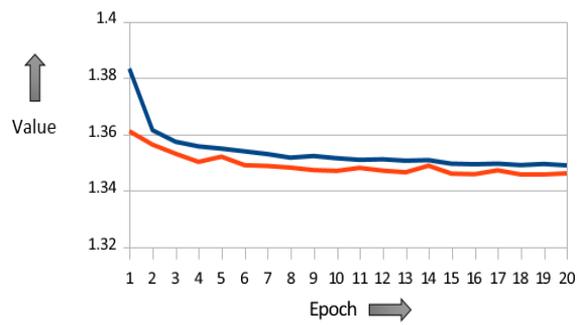
(d) Capsule network for full dataset

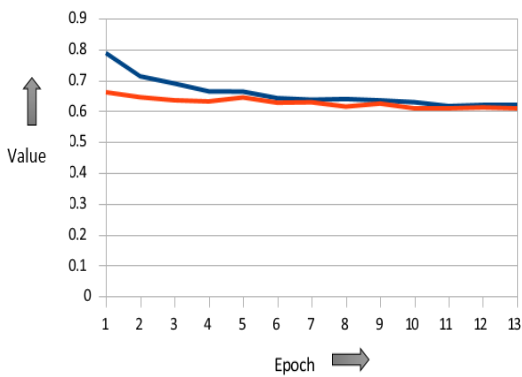
(e) Proposed VDSNet for sample dataset

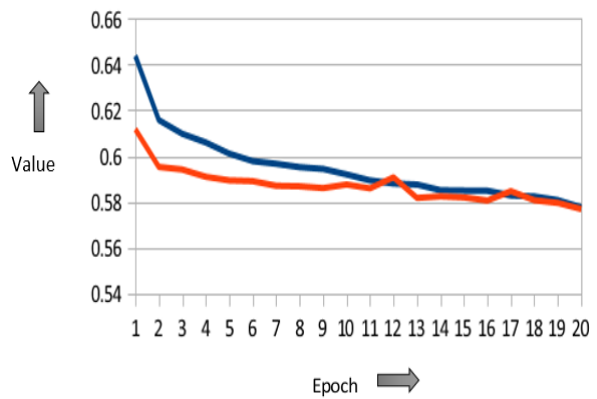
(f) Proposed VDSNet for full dataset

**Fig. 12.** Change of loss in training algorithms

From Fig. 12 it can be seen that the vanilla CNN exhibits the worst performance, it overfills too early and clogs because of the early stopping checkpoint model. Capsules network shows better performance than vanilla CNN, although the convergence is very slow. VDSNet performs the best but it converges very slowly, possibly owing to very little data on the features of the large images. Additional data in the full dataset may improve the convergence time.

We have found that the vanilla CNN stops and overfills by the model of early stopping of VDSNet. The convergence is fast as well as is still too useful convergence, will also have provided higher results if this paper train this model using more epoch. The performance of the capsule network is better than vanilla CNN, however it has slower convergence. From the plots of Fig. 12, it can be seen that VDSNet exhibits the best performance with some specific parameters as declared above.

**7.2 Justification**

Based on the accuracy of the approaches on the full dataset and the sample dataset, different models can be compared as shown in Table 1.

**Table 1.** Comparison of recall, precision, $F_\beta$ score, validation accuracy and training time for different models

| Dataset | Structural design | Recall | Precision | $F_\beta$ (0.5) score | Validation Accuracy | No. parameters | Training time (seconds) |
|---|---|---|---|---|---|---|---|
| Sample Dataset | Vanilla gray | 0.50 | 0.58 | 0.56 | 50.7% | 321225 | 2 |
| | Vanilla RGB | 0.59 | 0.62 | 0.61 | 51.8% | 322793 | 2 |
| | Hybrid CNN VGG | 0.56 | 0.65 | 0.63 | 68% | 15252133 | 16 |
| | **VDSNet** | **0.64** | **0.62** | **0.64** | **70.8%** | **15488051** | **19** |
| | Modified CapsNet | 0.42 | 0.71 | 0.45 | 59% | 12167424 | 37 |
| | Basic CapsNet | 0.60 | 0.62 | 0.62 | 57% | 14788864 | 75 |
| Full Dataset | Vanilla gray | 0.58 | 0.68 | 0.66 | 67.8% | 321225 | 51 |
| | Vanilla RGB | 0.61 | 0.68 | 0.66 | 69% | 322793 | 53 |
| | Hybrid CNN VGG | 0.62 | 0.68 | 0.67 | 69.5% | 15252133 | 384 |
| | **VDSNet** | **0.63** | **0.69** | **0.68** | **73%** | **15488051** | **431** |
| | Modified CapsNet | 0.48 | 0.61 | 0.58 | 63.8% | 12167424 | 856 |
| | Basic CapsNet | 0.51 | 0.64 | 0.61 | 60.5% | 14788864 | 1815 |

From Table 1, it can be seen that the best model is VDSNet which is better than the benchmark vanilla CNN. It can also be seen that the $F_{0.5}$ score of VDSNet is 0.68. The training time is greater than vanilla CNN. However, VDSNet model can be improved by continuing training with more epochs. On the other hand, the capsule network model does not seem to work well; the number of parameters is only equivalent to VDSNet, but the training time is much longer. VDSNet has $F_{0.5}$ score of 68% with 73% validation accuracy. It still does not meet the requirement to use in hospitals, need more time and computer power to further analyze the data, improving the algorithm can meet the requirements. However, this is also a good first step, and this result is very good when the normalized dataset is public and there are many mistakes in labeling.

There is a minimal scope of direct comparison with existing researcher works because the dataset used in this paper is entirely different and has several limitations compared to other datasets. Though it is not possible to make a direct comparison with the previous work. However, we have tried to make a comparison with some works. The work in [41] applied AlexNet, GoogLeNet, VGGNet-16 and ResNet-50 on eight common thoracic pathology classification using ChestX-ray8 database. But, we have not performed pathology localization accuracy using our model. So, no direct comparison is possible with [41]. Tang et. al. [42] achieved 62.7% AUC using U-Net autoencoder and 73.7% using U-Net autoencoder and discriminator for the classification of normal and abnormal lung conditions. They used general adversarial networks which is complex compared to our proposed method. We have achieved 73% validation accuracy and 74% AUC using VDSNet. Choudhary et al. [44] achieved 83.67% accuracy using their proposed CNN model having six layers. The achieved accuracy of the proposed VDSNet is less than that reported in [43]. In future, the accuracy of VDSNet can be increased by inserting additional layers. The implementation of multi-label chest X-Ray classification using the model in [44] will also be tried as future work. Different ResNet architectures are different from our benchmark model. It should be noted that Fibrosis can be found out from the chest X-ray image and can be evaluated in terms of confident score. So, it can play an important role in COVID-19 detection. Our future target will be to find out a suitable model from the reference paper [3, 41-44] in order to detect the lung diseases of COVID-19 affected patients.

## 7.3 Free-Form Visualization

In this research, we test with twenty random instances, the surgeon, either a patient or a physician, just completed records about age, X-rays, view position, and gender. We have evaluated and detected the illness of a patients before moving forward with the investigation on more significant trials. For the purpose of the prediction of diseases, we have calculated the $F_\beta$ score where $\beta$ is 0.5. It means that we are determined the condition of a patient such as the condition of sadness and shock before formal diagnosis. Most of the results are exactly the same (Fig. 13 (a, b)), but there are also some cases that are wrong (Fig. 13 (c)). The confident score for fibrosis finding case is 58.5842%. The confident score for Pneumothorax finding case is 48.33%.

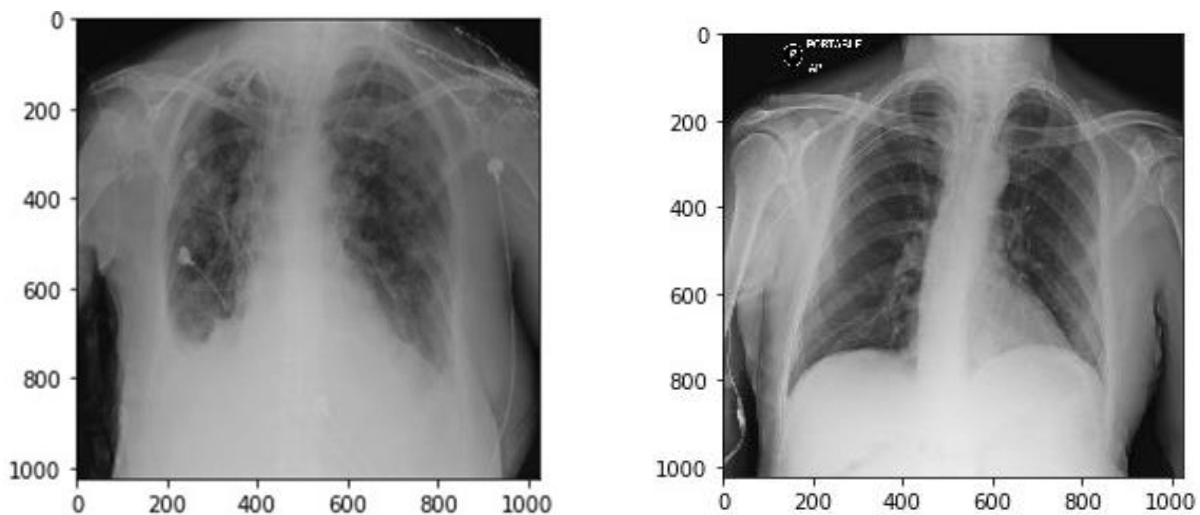

(a) Fibrosis finding case, confident score: 58.5842%

(b) No abnormality finding case, confident score: 7.4103%

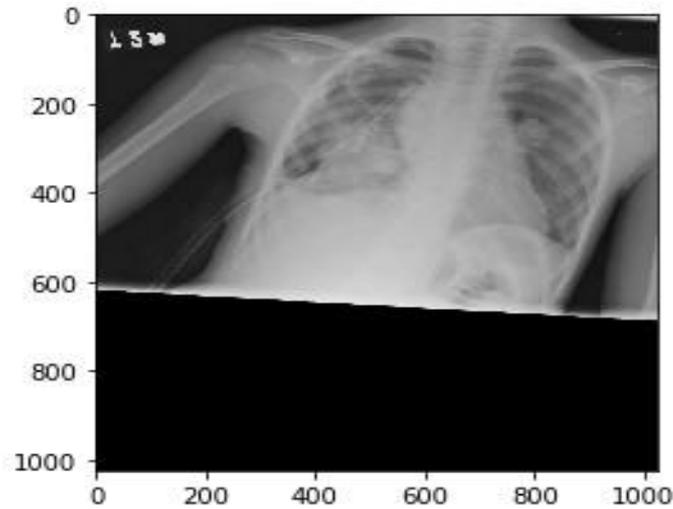

(c) Pneumothorax finding case, confident score: 48.3327%

**Fig.13.** Analytical lung image dataset with finding

There are some demerits of the prediction that the ill person is not ill, as the system ignores shocking patients as well as requires more tests, before the doctor provides the ultimate diagnosis. The β of the F score is 0.48 for the confidential cases which is proximate to the threshold. So, we have selected β as 0.5. It means that the chance of the illness is approximately half.

**7.4 Reflection**

We detect the lung disease using the patient's lung X-ray data and extra records. The ideal solution of this paper is to have a hybrid CNN with the description of the data process as follows:

(i) Research for support data, domain information, resolved issues, approaches, and solution data for similar paper. Some potential methods are investigated and listed.
(ii) Dataset of a sample data is downloaded with metric selection, preprocessing, and analyzed.
(iii) We have tested multiple structures, improved and tested on a sample lung dataset.
(iv) Finally, we have used best architects for the purpose of testing the full lung dataset, continued improving.

**7.5 Improvement**

As future work, this paper can be extended in a number of ways. Some of these are mentioned as follows.
(i) The model needs testing in order to differentiate each type of lung diseases. As a result, the data problem can be explained for each disease which is very skew.
(ii) The proposed model should be trained with a huge number of epochs with the change of a few parameters for getting fast convergence.
(iii) The probability of getting significant features will be increased if the size of training shots can be increased. But this can increase the training time.
(iv) Several pre-trained models can be experimented in order to implement CNN with the fusion of VGG.

(v) Very complex "locnet" module has been used in order to implement hybrid CNN with the addition of a spatial transformer.

(vi) In order to extract more features, CapsNet has been proposed after adding some more layers. However, it will lead to very long training time.

Moreover, VDSNet can be useful for other application areas [3, 45-47] as well. Particularly, VDSNet can be applied to X-ray images of suspected COVID-19 patients to predict whether patients have COVID-19 related pneumonia, or not [3].

## 8. Conclusion

In this work, a new hybrid deep learning framework termed as VDSNet is proposed for detecting lung diseases from X-ray images. The new model is applied to NIH chest X-ray image dataset collected from Kaggle repository. For the case of full dataset, VDSNet shows the best validation accuracy of 73%, while vanilla gray, vanilla RGB, hybrid CNN VGG, basic CapsNet and modified CapsNet have accuracy values of 67.8%, 69%, 69.5%, 60.5% and 63.8%, respectively. VDSNet exhibits a validation accuracy value of 73% which is better than the 70.8% accuracy value in case of sample dataset. On the other hand, VDSNet requires a training time of 431 seconds for the case of full dataset which is much higher than the 19 second time required for sample dataset.

In order to make the proposed VDSNet useful in hospitals, additional progresses are required to enhance the precision of the model. Generally, basic CNN has poor performance for rotated, tilted or other abnormal image orientation. Therefore, hybrid systems have been executed in order to improve the accuracy without increasing the training time. The results described in the paper recommend that the deep learning models can be utilized to improve the diagnosis compared to the traditional methods. As a result, the quality of the affected patient's treatment can be improved. Our hybrid approach can efficiently detect the inflammatory area in chest X-ray images. This research work faces some challenges at the time of handling the large scale dataset. Hence, the use of small datasets can provide good accuracy but it will not be effective in real applications. In future, we will apply modified VGG or other new transfer learning algorithms to the sample and full datasets and then make a hybrid algorithm with the fusion of GoogLeNet, AlexNet, and ResNet-152 architecture. We will also prepare a dataset by combining two or more chest X-ray datasets and then apply hybrid algorithms on the combined dataset for detecting various lung diseases. Future research scopes will also include the implementation of image data augmentation techniques such as color space augmentations, kernel filters, feature space augmentation, etc., in order to increase the accuracy in automated chest X-ray diagnosis system. In future, the proposed new VDSNet method can be applied to X-ray images of suspected COVID-19 patients in order to predict whether those patients have COVID-19 related pneumonia, or not.